\documentstyle[aps,prl,multicol,epsfig]{revtex}
\begin{document}
\draft
\title{Transition from damage to fragmentation in collision of solids}

\author{Ferenc Kun${}^{1,3}$\footnote{Email
address:feri@ica1.uni-stuttgart.de} and Hans J. Herrmann${}^{1,2}$}

\address{${}^1$Institute for Computer Applications (ICA1) \\
 University of Stuttgart, Pfaffenwaldring 27, D-70569 Stuttgart,
 Germany \\
${}^2$ PMMH, ESPCI, 10, rue Vauquelin, 75231 Paris Cedex 5, France \\
${}^3$Department of Theoretical Physics, Kossuth Lajos University, \\ 
P.O.Box: 5, H-4010 Debrecen, Hungary}

\date{\today}
 
\maketitle

\begin{abstract}
We investigate fracture and fragmentation of solids due to impact at
low energies using a
two-dimensional dynamical model of granular solids. 
Simulating collisions of two solid discs we show that, depending on the
initial energy, the outcome of a collision process can be classified 
into two states: a damaged and a fragmented state with a sharp
transition in between. We give numerical
evidence that the transition point between the two
states behaves as a critical point, and we discuss the possible
mechanism of the transition.
\end{abstract}
 

\begin{multicols}{2} 
\narrowtext

\section{Introduction}

Fragmentation, {\em i.e.} the breaking of particulate materials into
smaller 
pieces is a ubiquitous process that underlies many natural phenomena
and industrial processes. The length scales involved in it
range from the collisional evolution of asteroids through the scale of
geological phenomena down to the
breakup of heavy nuclei \cite{hans,frag_book}. In most of the 
realizations of fragmentation processes the energy is imparted to the
system by impact, 
{\em i.e.} typical experimental situations are
shooting a projectile into a solid block, free fall impact
with a massive plate and collision of particles of the same size
\cite{exp_tur,exp_colli1,exp_colli2,exp_colli3,exp_colli4,exp_bohr,exp_meibom,exp_inao,exp_kadono}.
The most striking observation about fragmentation  is 
that the size distribution of fragments shows power law behavior  
independent on the microscopic interactions
and on the relevant length scales, 
 {\em e.g.}
the charge distribution of small nuclei resulted from collisions
of heavy ions exhibits the same power law behavior as the size
distribution of asteroids
\cite{hans,frag_book,exp_tur,exp_colli1,exp_colli2,exp_colli3,exp_colli4,exp_bohr,exp_meibom,exp_inao,exp_kadono}.
Experiments revealed that the
power law behavior of fragment sizes is valid for a broad interval of 
the imparted energy
\cite{exp_tur,exp_colli1,exp_colli2,exp_colli3,exp_colli4,exp_bohr,exp_meibom,exp_inao,exp_kadono},
which was also 
reproduced by computer simulation of sophisticated microscopic models
\cite{ferenc1,ferenc2,potapov1,potapov2,thornton,benz}. 
The observation of power law size distributions without control
parameter initiated the idea of self-organized criticality
\cite{exp_bohr,exp_meibom,exp_inao,exp_kadono,automata,soc} in
fragmentation and gave rise to numerous studies to understand the
dynamic origin of the power law
\cite{ferenc1,ferenc2,potapov1,potapov2,thornton,benz,automata,soc,zhang,astrom1,astrom2,botet,engelman1,gonzalo}.
Hence, during the past years much information has been accumulated
about fragmentation processes in the imparted energy range where a power law
size distribution occurs but the limiting case of low energies is
still not explored.   

Beside the general interest in fragmentation processes one can also
mention other fields where  
fracture and fragmentation of solid particles due to impact play an
important role. It is well known that in the flow of granular
materials a large part of the kinetic energy of the grains is
dissipated in the vicinity of their contact zone
during the collisions. Beside the viscous and plastic effects, the
dissipation by 
damaging is also an important source of energy loss in the flow.
Collision of particles occurs also in the solar system in
planetary rings. In this case the energy dissipation due to impact
damage might also influence the large scale structure formation in the
rings \cite{ice}. On larger length scale in the solar system, the
so-called collisional 
evolution of asteroids due to subsequent collisions, and the formation
of rubble piles in the asteroid
belt is still a challenging problem \cite{exp_colli3}. Among
industrial applications  
the breakup of agglomerates in chemical processes can be mentioned.
Due to experimental difficulties, the computer simulation of microscopic
models is an indispensable tool in the study of these impact phenomena
\cite{ferenc1,ferenc2,potapov1,potapov2,thornton,benz}.

In the present paper we want to elaborate the
impact fracture and fragmentation of solids at low 
imparted energy using a
two-dimensional dynamical model of breakable granular solids. 
Simulating collisions of two solid discs we show that, depending on the
imparted energy, 
the outcome of a collision process can be classified
into two states: a damaged and a fragmented state with a sharp
transition in between. Analyzing the energetics of the impact and the
resulted fragment size distributions we give numerical
evidence that the transition point between the damaged and 
fragmented states behaves as a critical point. The transition 
proved to be the lower bound for the occurrence of power law size
distributions. The possible mechanism of the transition between the two
states is discussed. In spite of the specific features of the system 
studied here, most of our results can be considered generally valid for
impact phenomena mentioned above.

After giving a short summary of the main ingredients of our model in
Sec.\ II., the numerical results concerning to the energetics of the
collision process and to the size distribution of fragments will be
presented in Secs.\ III.\ and IV., respectively. In Sec.\ V.\ we
discuss the possible 
mechanism of the transition and  some general consequences
of our work for other types of fragmentation phenomena. 

\section{The model} 

Recently, we have worked out a two-dimensional dynamical model of
deformable, breakable granular solids, which enables us to perform
molecular dynamics simulation of  
fracture and fragmentation of solids in various experimental
situations \cite{ferenc1,ferenc2}. Our model is an extension of
those models which are used 
to study the behavior of granular materials applying randomly shaped
convex polygons to describe grains \cite{hjt}. To capture the elastic behavior
of solids we connect the
unbreakable, undeformable polygons (grains) by elastic beams. The
beams, modeling cohesive forces between grains, can be broken according to a
physical breaking 
rule, which takes into account the stretching and bending of the
connections. The breaking rule contains two parameters
$t_{\epsilon}, t_{\Theta}$ controlling the relative importance of the
stretching and bending breaking modes, respectively.  The energy
stored in a beam just before breaking is released in the breakage
giving rise to energy dissipation. 
The average value of the energy dissipated by the breakup
of one contact defines the crack surface energy $E_s$ in the model
solid as a function of the breaking parameters $t_{\epsilon}$, and
$t_{\Theta}$. At the broken beams along the surface of the polygons 
cracks are generated inside the solid and as a result of the successive
beam breaking the solid falls apart. The fragments are defined
as sets of polygons connected by remaining intact beams. In the
framework of this discrete model one can introduce the notion of the
binding energy $E_b$ as the energy  released in the case of complete
disintegration, {\em i.e.} in the case when all the fragments are single
polygons. Hence,
\begin{eqnarray}
  \label{binding}
  E_b = N_c \cdot E_s,
\end{eqnarray}
where $N_c$ denotes the total number of grain-grain contacts in the
solid. Note 
that $E_b$ is proportional to the volume of the sample and can be
considered as a natural energy scale of the system.

The time evolution
of the fragmenting solid is obtained by 
solving the equations of motion of the individual polygons until the
entire system 
relaxes, {\em i.e.} there is no breaking of the beams during some
hundreds consecutive time steps and there is no energy stored in
deformation.  
For more details of the model's definition see
Refs.\ \cite{ferenc1,ferenc2}. 

We have applied the model to study shock
fragmentation of solids in various experimental situations. Namely,
simulations were performed to study
the fragmentation of a solid disc caused by an explosion in the
middle \cite{ferenc1}, the breaking of a rectangular block due
to the impact with a projectile \cite{ferenc1}, and the collision of
two macroscopic bodies (discs) \cite{ferenc2}. The model proved to be
successful in reproducing the experimentally observed subtleties of
fragmenting systems, {\em e.g.} the power law mass
distribution of fragments was found independent on the initial
conditions, with an exponent in the vicinity of two, slightly depending
on the initial energy \cite{ferenc1,ferenc2}. 

\section{Damage and fragmentation} 

\begin{figure}
\begin{center}
\epsfig{bbllx=73,bblly=286,bburx=393,bbury=518,
file=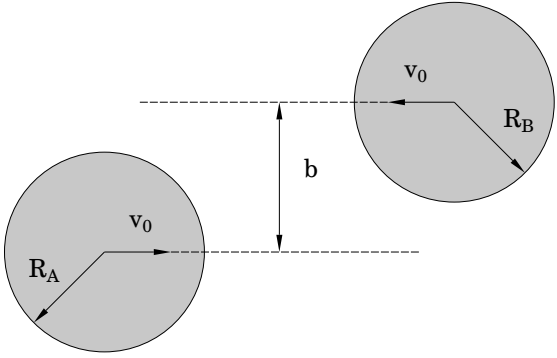,
  width=7.5cm}
 \caption{\small 
The schematic representation of the collision of two disks. In all the
simulations $R_A$=$R_B$ was used and the
value of $b$ was set to 0.}
\label{fig:fig1}
\end{center}
\end{figure}

In the present paper we apply our model 
to explore the properties of impact fragmentation processes at low 
imparted energy. 
For this purpose we carried out molecular
dynamics simulation 
of the collision of two solid discs of the same size, similar to
Ref. \cite{ferenc2}. The disc-shaped granular solid was obtained
starting from the Voronoi-tessellation of a square
\cite{moukarzel,lauritsen} and cutting out a
circular window in the middle. This way of construction gives rise to
a certain surface roughness of the particles. 
The schematic representation of the collision of two particles is
depicted in Fig.\ \ref{fig:fig1}.
For simplicity, in the present studies
only central collisions were considered, {\em i.e.} the value of the impact
parameter $b$ was set to zero in all the simulations, and only the size
of the particles $R$ and the surface
energy $E_s$ were varied. In the simulations the energy-release by beam
breaking is the only possible source of dissipation.
The energy of the collision $E_0$ is
defined as the total initial kinetic energy of the colliding bodies 
\begin{eqnarray}
E_0 = R^2\pi\rho v_0^2,
\end{eqnarray}
where $\rho$ is the mass density and $v_0$ denotes the
initial velocity of the particles. When 
$E_0$ is smaller 
than the so-called damage threshold $E_d$, the collision results 
simply in elastic rebound without internal damage (without breaking of
beams). To achieve damaging, $E_0$ has to surpass the damage threshold
$E_d$, which is determined by the surface energy $E_s$ of the model.

It is generally accepted that impact fragmentation phenomena exhibit
so-called energy scaling, {\em i.e.} the result of the fragmentation
process 
only depends on the value of the specific energy defined as the imparted
energy divided by the total mass of the system.
To characterize the collision events we introduce a dimensionless
parameter $\eta$ with the definition
\begin{eqnarray}
  \label{eta}
  \eta = \sqrt{\frac{E_0}{E_b}} \sim \frac{v_0}{\sqrt{E_s}}.
\end{eqnarray}
This choice of parameter $\eta$ has the advantage with respect to the
specific energy that besides the dependence on the global size it
includes also the specific material type through the surface energy $E_s$.
Since in the present study we focus on the evolution of damage with
increasing impact energy, 
simulations were performed at fixed values
of $R$ and $E_s$ varying  $\eta$ in the range $\eta_0 < \eta <
1.5$.  Here $\eta_0$ corresponds to the damage threshold $\eta_0 =
\sqrt{\frac{E_d}{E_b}}$. Our results concerning 
to impact 
fragmentation at $\eta > 1.5$ can be found in Ref.\ \cite{ferenc2}
including detailed analysis of the dynamics of the collision process
and comparison to experiments. The values of the most important
parameters of the present simulations are summarized in Table
\ref{table_0}. 

\begin{figure}
\begin{center}
\epsfig{bbllx=166,bblly=3,bburx=430,bbury=636,
file=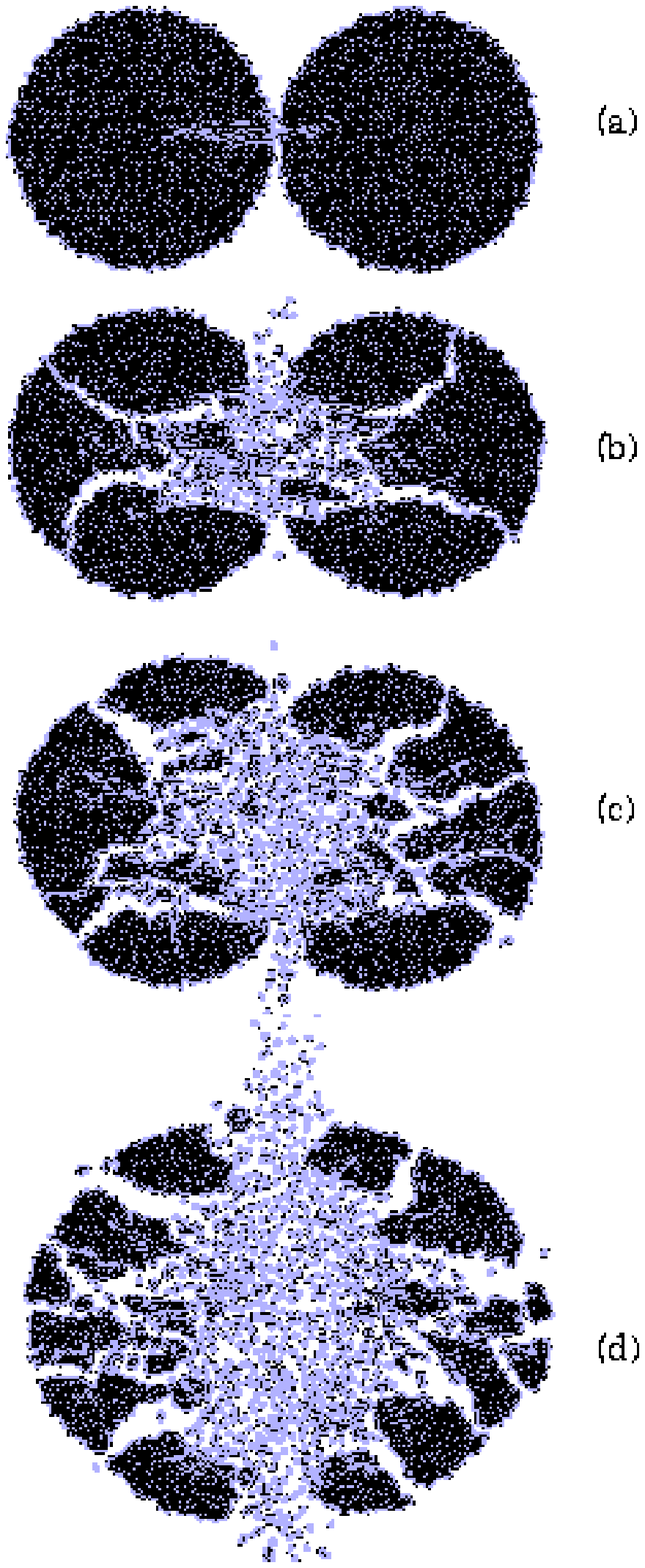,
  width=8cm}
 \caption{\small 
The final breaking scenarios of collisions of discs 
at different
impact energies $E_0$. The radius of the discs is $R=20 \mbox{cm}$,
while the average size of a randomly shaped polygon is $1
\mbox{cm}$. The lines connecting the center of mass of the 
neighboring polygons symbolize beams. The missing beams have already
been broken. The values of the parameter $\eta$ are 0.09, 0.2, 0.3,
0.5 for $a, b, c$ and $d$, respectively.}
\label{fig:fig2}
\end{center}
\end{figure}

In Fig.\ \ref{fig:fig2} we present the 
final breaking scenarios (relaxed states) of the collision
process obtained by simulations at four different values of $\eta$.
It can be observed that for $\eta \sim \eta_0$ (Fig.\ $\ref{fig:fig2}a$)
cracking by beam breaking
mainly occurs in the vicinity of the contact surface of the
two bodies and the bulk remains practically intact. 
Since this gentle collision does not cause size reduction of the
two particles,  
this case can be
considered as an inelastic impact of discs, where energy dissipation is
solely due to 
internal damage.
Increasing the impact energy $E_0$, {\em i.e.} increasing $\eta$,
gives rise to more broken beams in Fig.\ $\ref{fig:fig2}b$ and the
solids break 
into pieces. Around the impact site of the bodies the fragments are smaller
(single polygons, some pairs and triplets), and there are a few much larger
fragments the size of which is still comparable to the original size
of the bodies. 
Further increase of $\eta$ (Fig.\ $\ref{fig:fig2}c, d)$ mainly
results in breakup of 
the large fragments into smaller ones giving rise to fragmentation of
the entire solids. 
Based on the above qualitative picture, we classify the
outcome of a collision into two states, {\em i.e.} {\it damaged} and {\it
  fragmented} states  distinguished by the size of the largest
remaining piece. The colliding bodies are considered to be {\it damaged}
when the size of the largest piece is comparable to the original size
of the bodies (see Fig.\ $\ref{fig:fig2}a, b)$, while the {\it fragmented}
state is characterized by the absence of such large pieces (Fig.\
$\ref{fig:fig2}c, d)$. 
In the following we point out that the above qualitative picture is
also supported by the quantitative 
analyzis of the energetics of the collision process
and of the behaviour of the resulting size distribution of fragments.
Namely, evaluating the energy released by the breaking of beams, and
quantities related to the size distribution of fragments we will 
justify that the behaviour of the colliding system has two
different regimes with a sharp transition in between.

\begin{figure}
\begin{center}
\epsfig{bbllx=150,bblly=423,bburx=483,bbury=728,
file=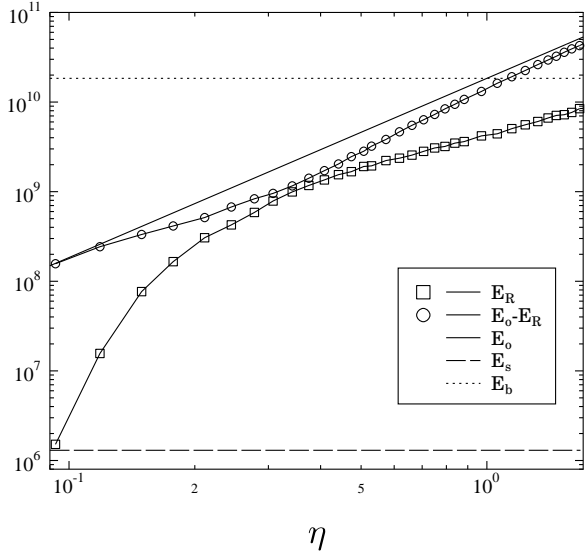,
  width=8.5cm}
\caption{\small 
The energy released by beam breaking $E_R$ and the kinetic energy
stored in the motion of fragments $E_0 - E_R$. For
comparison, the 
surface energy $E_s$ of the model, the binding energy of the sample
$E_b$ and the energy of the impact $E_0$ are also indicated.   
The energy unit is {\em erg}. For the exact values see Table \ref{table_0}.}
\label{fig:fig3}
\end{center}
\end{figure}

\begin{figure}
\begin{center}
\epsfig{bbllx=138,bblly=433,bburx=483,bbury=728,
file=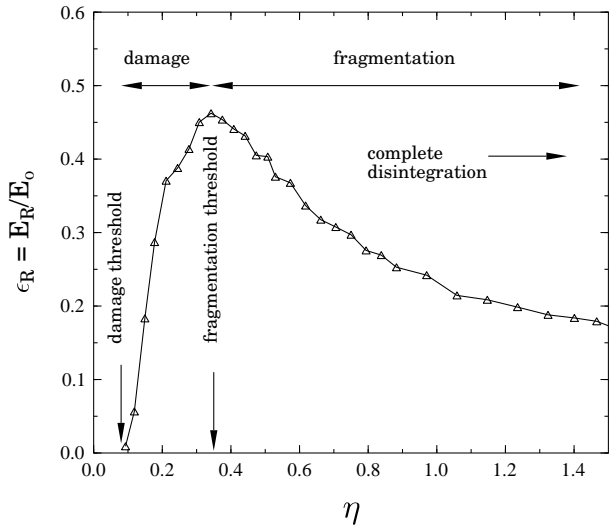,
  width=8.5cm}
\caption{\small 
The ratio $\epsilon_R$ of the energy released by breaking $E_R$ and
the total kinetic energy $E_o$. The transition
point (fragmentation threshold) between the damaged and fragmented
states is identified with the position of the maximum of
$\epsilon_R$. 
}
\label{fig:fig4}
\end{center}
\end{figure}

The energetics of the collision process corresponding to the samples of
Fig.\ \ref{fig:fig2} is summarized  on
Figs.\ \ref{fig:fig3}, and \ref{fig:fig4}. Fig.\ \ref{fig:fig3} shows the
energy $E_R$  released  by beam breaking, together with the remaining kinetic
energy $E_0-E_R$ stored in the motion of fragments as a function of
$\eta$. For the purpose of 
comparison, the 
surface energy $E_s$ of the model, the binding energy of the sample
$E_b$ and the energy of the impact $E_0$ are also plotted in the
figure. When the impact energy $E_0$ surpasses the damage threshold
$E_d$ ($\eta$ surpasses $\eta_0$) the released energy $E_R$ takes its
smallest value, which is 
equal to the surface energy $E_s$ of the model (only one beam is
broken). The largest possible 
value of $E_R$ is equal to the binding energy of the sample $E_b$,
which can only be reached in the case of complete disintegration. According
to the simulations, this limiting case is hard to achieve, since on the
surface of the colliding discs opposite to the impact site, small
fragments comprising a few polygons can always escape. It can also be
seen in the figure that the
impact energy $E_0$ producing complete disintegration has to be much
larger than the binding energy $E_b$. Note that similar qualitative
behaviour was found   simulating free fall impact
of a disc with a massive plate \cite{thornton}.

The most
interesting observation of Fig.\ \ref{fig:fig3} is that the curve of
$E_R$ versus\ $\eta$ is composed of two distinct parts, {\em i.e.} a
rapidly increasing 
low energy part, and a slowly increasing high energy part.
These two regimes are separated by a point where $E_R$ is practically equal
to the remaining kinetic energy $E_0-E_R$, which implies that at this
point half of 
the total kinetic energy $E_0$ is released by cracking. 
The two regimes can be better observed in Fig.\
\ref{fig:fig4} 
where the ratio $\epsilon_R$ of $E_R$ and $E_0$ is plotted against $\eta$. 
The curve of $\epsilon_R$ has a maximum, the value of which is in the
vicinity of 0.5. This maximum separates a rapidly increasing and a
slowly decreasing part, which we associate to the damaged and
fragmented regimes introduced above, respectively. We identify the
transition point 
between the two states with the position of the maximum
called {\it fragmentation threshold} $\eta_c$.  

\begin{figure}
\begin{center}
\epsfig{bbllx=148,bblly=399,bburx=481,bbury=701,
file=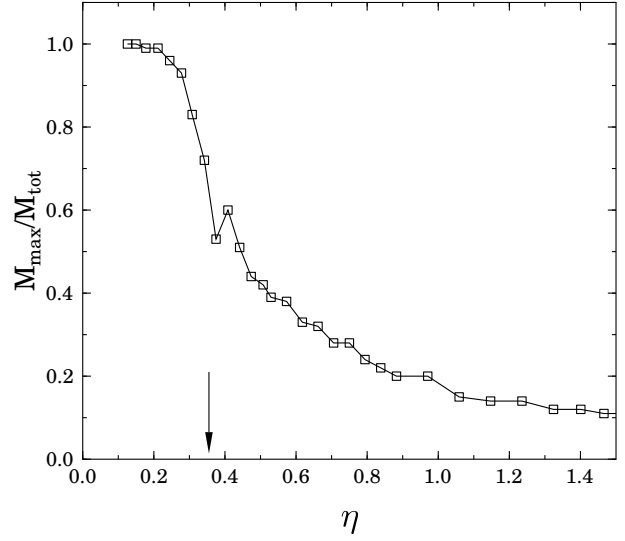,
  width=8.5cm}
\caption{\small 
The mass of the largest fragments divided by the total mass of the
colliding system. The arrow indicates the fragmentation threshold
$\eta_c$. }
\label{fig:fig5}
\end{center}
\end{figure}

The separation of the two regimes and the identification of the transition
point is also supported by the behaviour of 
the mass of the largest fragment as a function of $\eta$.
In Fig.\ \ref{fig:fig5} we present the sum of the mass of the largest
fragments  of the two bodies $M_{max}=M_{max}^A+M_{max}^B$ normalized
by the total mass  $M_{tot} = M^A+M^B$ versus \ $\eta$. The monotonically
decreasing function has a distinct curvature change, the position of
which $\eta_c = 0.36$ coincides with the
maximum of the energy release curve $\epsilon_R$. 
This also 
confirms that the behaviour of the system sharply changes at a
specific value of $\eta$ that we called 
fragmentation threshold $\eta_c$.

\begin{figure}
\begin{center}
\epsfig{bbllx=128,bblly=399,bburx=472,bbury=701,
file=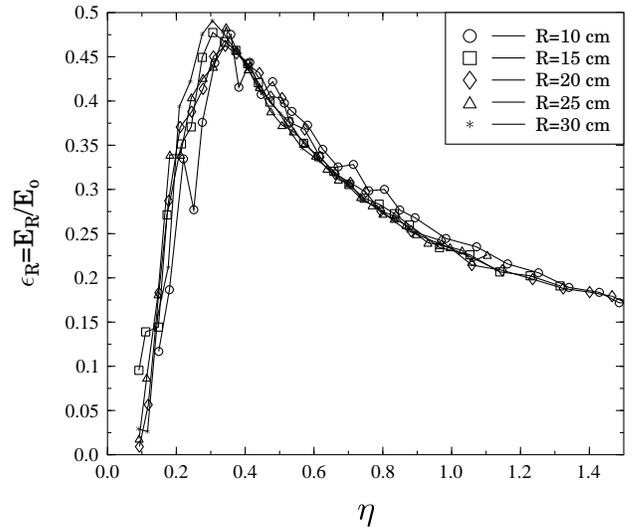,
  width=8.5cm}
\caption{\small 
The energy release curve $\epsilon_R$ at different system sizes $R$.
}
\label{fig:fig6}
\end{center}
\end{figure}

To clarify the influence of the overall size of the colliding bodies
on the behaviour of the fragmenting system, simulations were performed
varying the radius $R$ of the particles between 10 and 30 cm.  Note
that the average size of the randomly shaped unbreakable polygons is
1cm. Fig.\ \ref{fig:fig6} shows $\epsilon_R$ as a function of $\eta$ for
several values of $R$. It can be observed that the qualitative behaviour of
the curves is universal, it is independent  on the system size, but
the position of the  transition point $\eta_c$ is slighly shifted
downward with increasing $R$. 

\section{The size distribution of fragments}

To reveal the nature of the transition from the damaged to the
fragmented state, the evolution of the fragment size
distribution with varying impact energy is a crucial point. Since the
dissipated energy $E_R$ is proportional to the total surface of
cracks it is
expected that the two regimes should clearly show up also
in the type of disintegration of the solids, {\em i.e.}
in the behaviour of the size distribution of fragments.
In the present chapter we demonstrate that in the
vicinity of $\eta_c$ the behaviour of the fragment
size distribution shows strong similarities to that of the size
distribution of clusters in systems undergoing second order phase
transition. As an example, one can mention the behaviour of the cluster
size distribution in percolation around 
the critical point $p_c$,  or the size distribution of droplets in the
vicinity of the critical temperature $T_c$ in the
case of the liquid-gas phase transition. Throughout this chapter we refer
to percolation on finite lattices for the purpose of comparison.

\begin{figure}
\begin{center}
\epsfig{bbllx=134,bblly=105,bburx=481,bbury=642,
file=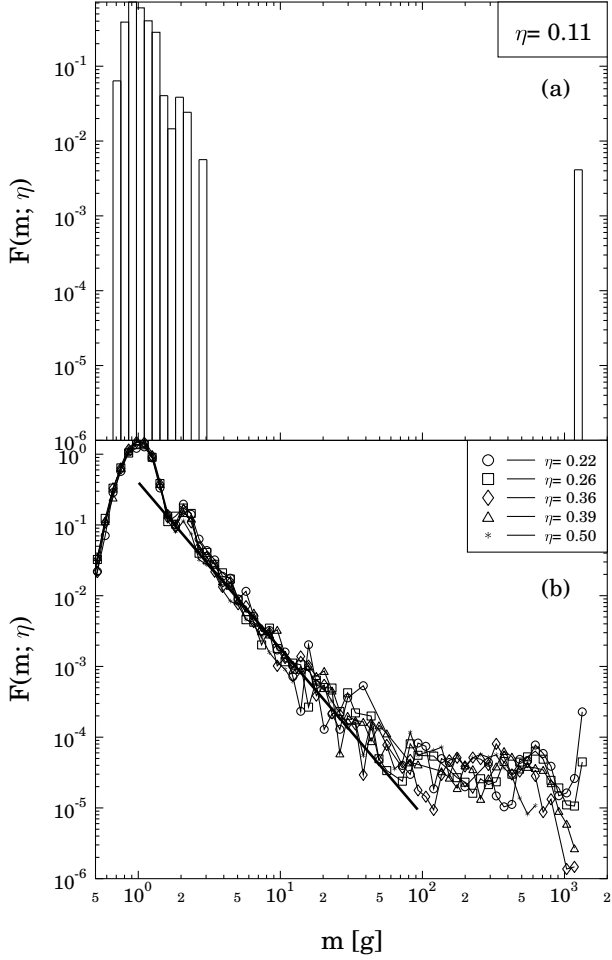,
  width=8.5cm}
\caption{\small 
Mass distribution of fragments at different values of $\eta$ below and
above the fragmentation threshold $\eta_c$. The peak of small
fragments is due to the fact that single polygons are unbreakable in the
model. The mass of the colliding discs with radius $R=20$cm was 1250
gram.} 
\label{fig:fig7}
\end{center}
\end{figure}

The fragment mass histograms $F(m;\eta)$ corresponding to the colliding system
of the preceding chapter are shown in Figs.\
$\ref{fig:fig7}a,b$ for several values of $\eta$ below and
above the transition point $\eta_c$.
Here $F(m;\eta)$ denotes the number of fragments with mass $m$ divided by
the total 
number of fragments, averaged over 10 collision events having the same
$\eta$. In order to obtain the same statistics of the 
distribution at all sizes a 
logarithmic bining was used, {\em i.e.} the bining is equidistant on
logarithmic scale. The histograms have two cutoffs. The lower one is due
to the existence of single unbreakable polygons while the upper one is
determined by the finite size of the bodies. 

In Fig.\ \ref{fig:fig2} we have shown that a collision at $\eta$ just
above $\eta_0$ results in only a few small fragments and two big ones
(with almost the same
size) but no fragments are generated in the intermediate mass
region. Hence, in Fig.\ $\ref{fig:fig7}a$ 
the corresponding mass distribution $F(m;\eta)$ has two peaks at small and
at large mass with a large gap in between. 
Increasing $\eta$ in Fig.\ $\ref{fig:fig7}b$, the main difference
compared to Fig.\ $\ref{fig:fig7}a$ is that the peak of the large
fragments gradually disappears, the large pieces break into
smaller ones giving rise to fragments in the intermediate mass region.
For $\eta \rightarrow \eta_c$ the mass distribution $F(m;\eta)$ in the
intermediate region tends to a power law:
\begin{eqnarray}
  \label{power}
  F(m) \sim m^{-\tau}.
\end{eqnarray}
The value of the exponent obtained at $\eta = \eta_c$ is $\tau = 2.27
\pm 0.05$. Our former simulations showed (see Ref.\ \cite{ferenc2})
that the power law behavior of $F(m;\eta)$
remains valid for $\eta > 
\eta_c$, but the value of the exponent
$\tau$ is a decreasing function of $\eta$,  in agreement
with most of the experimental observations 
\cite{exp_colli1,exp_colli2,exp_colli3,exp_colli4}.

\begin{figure}
\begin{center}
\epsfig{bbllx=136,bblly=341,bburx=475,bbury=643,
file=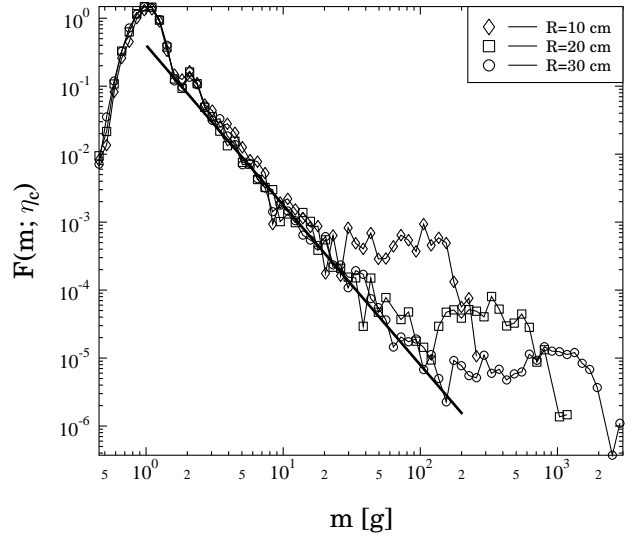,
  width=8.5cm}
\caption{\small 
Mass distribution of fragments at the transition point $\eta_c$ for three
different values of the system size $R$.}
\label{fig:fig8}
\end{center}
\end{figure}

To clarify the effect of the overall size of the colliding bodies $R$
on the shape of the mass distribution $F(m;\eta)$ and on the value of the
exponent $\tau$, in Fig.\
\ref{fig:fig8} we plotted the fragment size distributions at the
transition point $\eta_c$ for three different values of $R$.
It can be seen that increasing $R$ the power law region of
$F(m;\eta_c)$ gets wider and the relative height of the hump at large
fragments decreases, 
but the value of the exponent $\tau$ does not change. It is important
to note that in the 
case of percolation on finite lattices the size
distribution of clusters  shows the same shape and 
dependence on the system size at the critical point \cite{stauffer}.  

More insight into the evolution of the shape of
$F(m;\eta)$ can be obtained by studying the moments of the
distribution as a function of $\eta$. 
The $k$th moment $M_k(\eta)$ of the histogram $F(m;\eta)$ is defined as
\begin{eqnarray}
  \label{moments}
  M_k(\eta) = \sum_m  m^k \cdot F(m;\eta).
\end{eqnarray}
In the case of critical phenomena like percolation or the liquid-gas
transition the moments $M_k$ of the cluster size distribution with $k > 1$ 
diverge at the critical point in the thermodynamic limit
\begin{eqnarray}
  \label{diverge}
  M_k \sim |\epsilon|^{-\mu_k},
\end{eqnarray}
where $\epsilon$ denotes the distance from the critical point,
{\em i.e.} $\epsilon = p - p_c$ for percolation or $\epsilon = T - T_c$ for
the liquid-gas transition.
In a finite system the moments $M_k$ have a finite maximum at the
transition point.  
Assuming gap scaling for $F(m)$ \cite{stauffer}
\begin{eqnarray}
  \label{gap}
  F(m) \sim m^{-\tau} f(m^{\sigma}\epsilon)
\end{eqnarray}
the moment exponents $\mu_k$ can be expressed in terms of $\tau$ and
$\sigma$
\begin{eqnarray}
  \label{muk}
  \mu_k = \frac{1+k-\tau}{\sigma},
\end{eqnarray}
and the behavior of the system in the vicinity of the transition
point can be characterized by only two independent exponents $\tau$ and
$\sigma$. 

\begin{figure}
\begin{center}
\epsfig{bbllx=137,bblly=371,bburx=472,bbury=671,
file=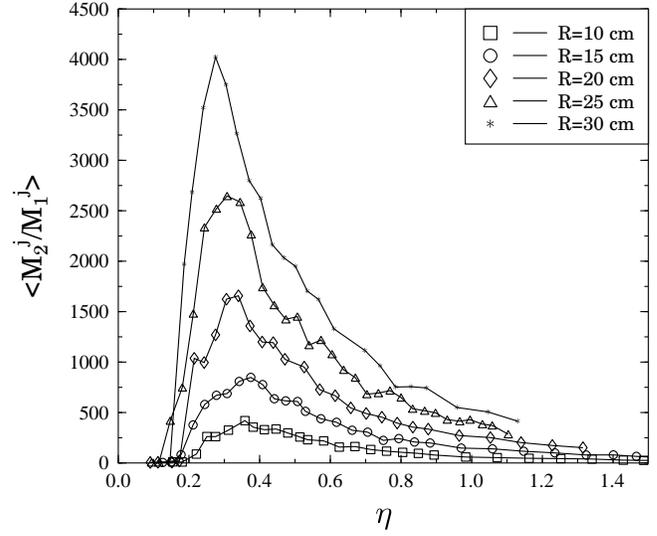,
  width=8.5cm}
\caption{\small 
$\left< M_2^j/M_1^j \right>$ as a function of $\eta$ for five different
values of $R$.
}
\label{fig:fig9}
\end{center}
\end{figure}

To test whether our small systems exhibit some trace of this
behavior we evaluated so-called {\it single event moments} $M_k^j$
defined as 
\begin{eqnarray}
  \label{singmom}
    M_k^j = \sum_m {}^{'} m^k \cdot n^j(m),
\end{eqnarray}
where the upper script $j$ refers to the $j$th collision event, 
$n^j(m)$ denotes the number of fragments with mass $m$
in event $j$ and
the prime indicates that the sum runs over all the fragments
excluding the largest ones of each of the two colliding bodies. The
study of single event moments 
was first suggested by X.\ Campi to reveal properties of fragmentation
processes with varying imparted energy \cite{campi1,campi2,campi3}. As
it was demonstrated in Refs.\ \cite{campi1,campi2,campi3}, the usage of single
event moments defined by Eq.\ (\ref{singmom}) instead of 
Eq.\ (\ref{moments}) has the advantage that part of the analysis, 
{\em e.g.} the study of the correlation of moments in scatter plots,
can be carried out without ordering the collision events according to
a parameter like in our case $\eta$.
Following Campi's ideas, we evaluated the ratio of $M_2^j$ and
$M_1^j$
\begin{eqnarray}
  \label{m2m1}
  \frac{M_2^j}{M_1^j} = \frac{\sum_m^{'} m^2\cdot
    n^j(m)}{\sum_m^{'} m\cdot  
    n^j(m)} = \overline{M}^j, 
\end{eqnarray}
which is equal to the average fragment
size $\overline{M}^j$ in event $j$. In Fig.\
\ref{fig:fig9} $\left< M_2^j/M_1^j \right>$ is plotted as a
  function of $\eta$ for several different values 
of the system size $R$. The brackets $<\dots>$ denote that each data
point was obtained as an average over 10 events having the same
$\eta$. One can observe that $\left< M_2^j/M_1^j \right>$ has a  
distinct maximum the position of which coincides  with the maximum of
the energy release curve $\epsilon_R$ in Fig.\ \ref{fig:fig6}, within
the precision of the calculations. It is very
important to note that 
increasing the system size $R$ the peak of $\left< M_2^j/M_1^j
\right>$ gets sharper, 
{\em i.e.} the height of the peak increases while the width of the
peak decreases, which is also typical for a critical point like the
percolation threshold occurring in finite size systems
\cite{stauffer,campi1,bauer}.   

\begin{figure}
\begin{center}
\epsfig{bbllx=233,bblly=314,bburx=460,bbury=689,
file=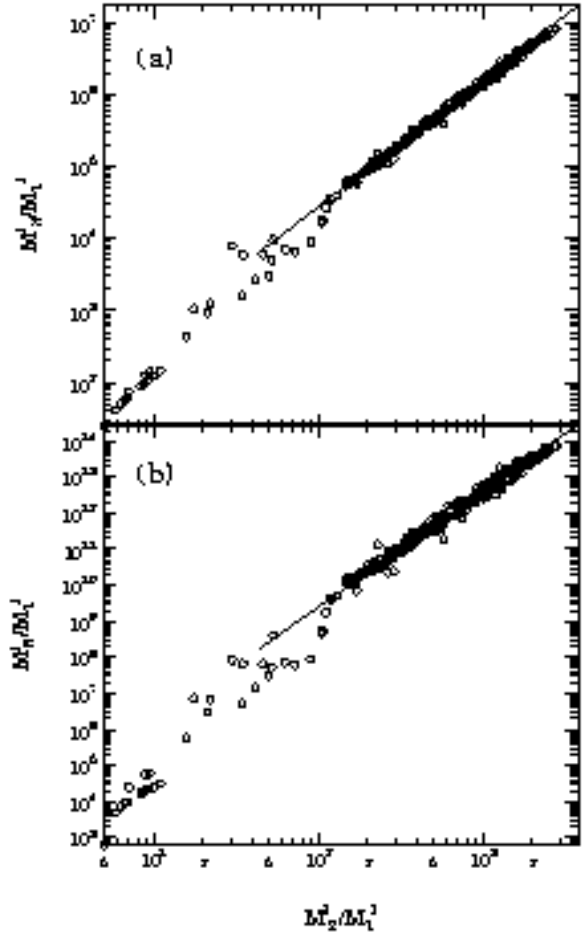,
  width=8cm}
\caption{\small 
Scatter plot of moments for the system of size R=20cm. The straight
lines were fitted to the data points in the range $M_2^j/M_1^j >
300$.} 
\label{fig:fig10}
\end{center}
\end{figure}

The validity of gap scaling Eq.\ (\ref{gap}) for the mass distribution
$F(m;\eta)$ can be easily tested by making scatter plots of moments $M_k^j$,
{\em i.e.} by plotting $M_k^j$ against $M_{k^{'}}^j$ with $k, k^{'} > 1$,
then checking the correlation of the moments and the validity of the
relation of the exponents Eq.\ (\ref{muk}). The scatter plot of two
pairs of moments is shown in Fig.\ \ref{fig:fig10}. For the
purpose of normalization the moments with $k > 1$ are divided by
$M_1^j$. 
In Fig.\ \ref{fig:fig10}
the collision events close to the transition point $\eta_c$ are
represented by points with the largest values of the moments. Events
with  $\eta$ far below or far above $\eta_c$ result in
smaller values of $M_k^j$, hence, the corresponding points in Fig.\
\ref{fig:fig10} fall closer to the origin. One of the most remarkable
features of Fig.\ \ref{fig:fig10} is that the moments are strongly
correlated in the region corresponding to the vicinity of the
transition point $\eta_c$ and a power law seems to be a reasonable fit
to the scattered points. By this fitting procedure one obtains
numerically the ratio of the moment exponents $\mu_k$ and
$\mu_{k^{'}}$ since
\begin{eqnarray}
  \label{scatt}
  M_k \sim M_{k^{'}}^{\mu_k/\mu_{k^{'}}}, \ \ \ \qquad \ \ \ k, k^{'}
  > 1.
\end{eqnarray}
The value of the exponents obtained as the slope of the straight lines
in Fig.\ \ref{fig:fig10} are $\mu_3/\mu_2 = 2.28 \pm 0.1$ and
$\mu_5/\mu_2 = 4.95 \pm 0.17$, which are in agreement with the
corresponding predictions 2.36 and 5.11 substituting $\tau =
2.27$ into Eq.\ (\ref{muk}). 

\begin{figure}
\begin{center}
\epsfig{bbllx=136,bblly=395,bburx=475,bbury=699,
file=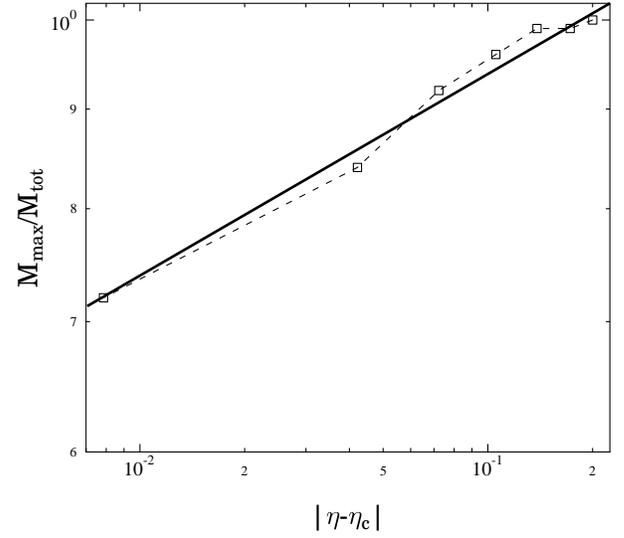,
  width=8.5cm}
\caption{\small 
Determination of the order parameter exponent $\beta$. The data of
Fig.\ {\protect \ref{fig:fig5}} is plotted here as a function of
$|\eta-\eta_c|$ for $\eta < \eta_c$. The exponent $\beta$ can be
obtained as the slope of the least square fitted straight line.}
\label{fig:fig11}
\end{center}
\end{figure}

\begin{figure}
\begin{center}
\epsfig{bbllx=136,bblly=395,bburx=475,bbury=699,
file=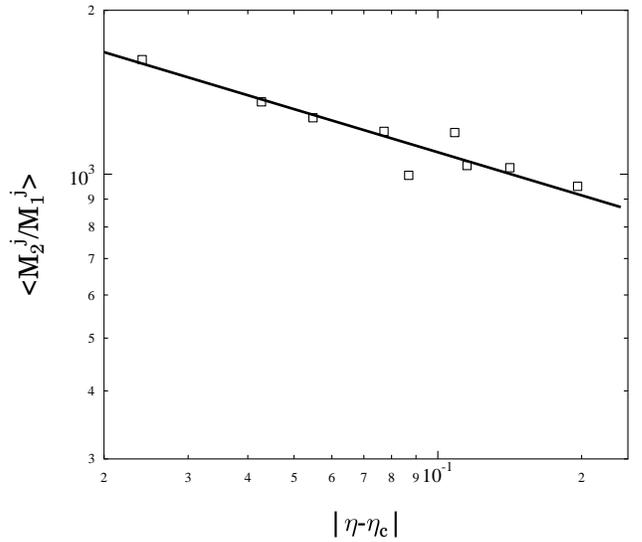,
  width=8.5cm}
\caption{\small 
To obtain the exponent $\gamma$ we plotted $<M_2^j/M_1^j>$ belonging to the
system size $R=20 \mbox{cm}$ of 
Fig.\ {\protect \ref{fig:fig9}} as a function of
$|\eta-\eta_c|$ for $\eta > \eta_c$. Data
points are taken only from the vicinity of $\eta_c$. The
exponent $\gamma$ is
obtained as the slope of the least square fitted straight line.}
\label{fig:fig12}
\end{center}
\end{figure}

Hence, we conclude that, within the limited
precision available in our small system, the mass distribution of
fragments fulfills the gap scaling relation Eq.\ (\ref{gap}), and the
damage-fragmentation transition occurs as a continuous phase
transition. The control parameter of the transition can be identified
with the parameter $\eta$, and the quantity characterizing the size
reduction $M_{max}/M_{tot}$ can be considered as the order
parameter. To complete the characterization of our system 
around $\eta_c$ we determined numerically the value of the order
parameter exponent $\beta$ defined as \cite{gold}
\begin{eqnarray}
  \label{beta}
  \frac{M_{max}}{M_{tot}} \sim |\eta - \eta_c|^{\beta}, \qquad \qquad
  \eta < \eta_c,
\end{eqnarray}
and the value of the exponent $\gamma$ characterizing the average
fragment size
\begin{eqnarray}
  \label{gamma}
  \left<\frac{M_2^j}{M_1^j}\right> \sim |\eta - \eta_c|^{-\gamma}.
\end{eqnarray}
The power law behavior given by Eq.\ (\ref{gamma}) is valid on both
sides of the critical point $\eta_c$.  (Note that $\gamma = \mu_2$.)
Using the data presented in Fig.\ \ref{fig:fig5} we plotted
$M_{max}/M_{tot}$ versus $|\eta - \eta_c|$ for $\eta < \eta_c$ in
Fig.\ \ref{fig:fig11}. It can be seen that a power law with an
exponent $\beta = 0.11 \pm 0.02$ gives a reasonable fit to the
data. In Fig.\ \ref{fig:fig12} $\left<M_2^j/M_1^j\right>$ belonging to
the system size $R=20 \mbox{cm}$ of Fig.\ \ref{fig:fig9} is depicted
as a function of  $|\eta - \eta_c|$ for $\eta > \eta_c$. Again a power
law with an exponent $\gamma = 0.26 
\pm 0.02$ can be fitted with a reasonable quality.
Since only two of the critical exponents are independent, we can check
the consistency of our description by pairing the exponents $\tau,
\beta, \mbox{and} \ \gamma$ and using the well known scaling laws
\cite{stauffer,gold} of critical phenomena to obtain the value of the
other exponents. It can be observed in Table \ref{table_1} 
that the numerical values of the exponents obtained by starting from
the three possible pairs agree well with each other within the error bars,
which gives further justification of our phase transition picture.

\section{Discussion}
In the present paper we applied our dynamical model of granular solids
to study impact fragmentation phenomena at low
values of the imparted energy. 
Analyzing the energetics of the fragmentation process and the resulting
size distribution of fragments we identified two distinct final states
of the impact process, {\em i.e.} {\em damaged} and 
{\em fragmented} states with a sharp transition in between. With a
detailed study of the behavior of the fragment mass distribution in
the vicinity of the transition point and its dependence on the finite
particle size we gave numerical evidence that
the transition point behaves as a critical point and the
{\em damage-fragmentation} transition occurs as a continuous phase
transition. The control parameter of the transition was chosen to be
the dimensionless ratio $\eta$ of the energy of impact and the binding energy
of the sample, and the order parameter was associated to the mass of
the largest fragment divided by the total mass. 

Based on the time evolution of the collision process obtained by the
simulations, the following qualitative picture of the mechanism of the
transition can be established: 
when two macroscopic solids collide damage first occurs in the surroundings
of the contact zone of the bodies. After this contact damage a
compressive elastic pulse expands radially through the bodies the
amplitude of which strongly depends on the amount of primary damage
occurred in the contact zone. The pulse reflects back from the free
boundary and after reflection tensile forces 
arise in the solids resulting in cracks in the bulk. The amount of
breaking in the bulk is mainly determined by the amplitude of the initial
pulse remaining after contact damage. 
The fragmentation of the colliding solids occurs above a specific value of
the impact energy where the amplitude of the remaining pulse will be
sufficient to give rise to the complete break up of the bulk into
pieces. At this specific energy value the behavior of the colliding
system sharply changes as we demonstrated by analyzing the energetics
and the size distribution of fragments.

 The idea of the existence of a so-called fragmentation phase
transition was first addressed in the field of nuclear physics where
the fragmentation of heavy nuclei due to impact is extensively
studied (for a recent review see Ref.\ \cite{traut}). Using
percolation based ideas it was shown that the  
disassembly of excited nuclei possesses a continuous phase transition
when the imparted energy is varied
\cite{campi1,campi2,campi3,bauer}. The order parameter of the
transition was associated to the size (charge) of the largest fragment.

Recently, the penetration of a steel ball into a solid plate with
varying impact energy was studied experimentally \cite{englman}. It
was reported that the so-called ballistic limit, {\em i.e.} the impact
energy where the perforation of the plate occurs, behaves as a
critical point, and the perforation occurs as a continuous phase
transition the order parameter of which was chosen to be the mass of
ejecta expelled from behind the target. It was argued that the
mechanism of the phase transition 
is the coalescence and percolation of randomly nucleated microcracks. 

In the analysis of the numerical results obtained by the computer
simulation of our dynamical model, we referred to the theory of
percolation on finite lattices just as an example of critical
phenomena occurring in finite systems, without assuming percolation of
microcracks. In this sense our analysis is free of model assumptions. 
In the snapshots of the final scenarios of collision
processes it can be seen that the complicated elastic field arising due to
propagation and interference of elastic waves gives rise to correlated
crack growth. Hence, in our case the assumption of 
percolation of cracks cannot be valid, and the values of the exponents
obtained numerically are different from the corresponding exponents of
percolation. 

An important feature of our model which can affect the result of a
fragmentation process is the existence of elementary,
non-breakable polygons, which hinders us from observing fragmentation on
lower scales. In Ref.\
\cite{potapov2}, using a discrete model of solids similar to ours,
detailed tests of the effect of the polygon size on the shape of the
mass distribution of fragments in impact fragmentation were performed.
It was found that if the tessellation of a solid is fine enough to
reproduce the macroscopic elastic behavior correctly, the effect of
the existence of unbreakable polygons on the size distribution of
fragments occurs solely in the vicinity of
the cutoff size, and the overall shape of the
distribution is not affected. This also implies that the phase
transition behavior 
revealed by our simulations is not influenced globally by the element
size, but it can be expected that the critical regime gets narrower
when the element size decreases at a fixed value of the macroscopic
size of the colliding bodies.

Studying kinetic models of fragmentation phenomena in Refs.\
\cite{botet,redner} it was pointed out that under certain circumstances
a so-called shattering transition might occur, when a macroscopic
amount of mass of the fragmenting system is transformed into a dusty
phase, {\em i.e.} into single polygons in our case. Since in discrete
dynamical models such a shattering transition can occur only at very
high impact energies its investigation is not included in the present
study. 

The binary break-up of bond percolation clusters due to the removal of
a single bond has also been studied in the context of fragmentation
phenomena. The characteristic quantities describing the fragmentation
of percolation clusters are the number of fragmenting bonds
$a_s(p)$, {\em i.e.} the number of 
those bonds in the cluster of size $s$ whose removal results in two
disconnected clusters, and the
probability $P_{s',s}(p)$ that the break-up of a cluster of
size $s$ results in a daughter cluster of size $s'$ \cite{fragperc}.
$a_s(p)$ and $P_{s',s}(p)$ are of fundamental interest because
they describe the connectivity of percolation clusters and 
they may also serve as an 
input for the rate equations describing fragmentation processes
as a sequence of binary events \cite{botet,redner,fragperc}. 
At the critical point $p=p_c$ both $a_s(p_c)$ and
$P_{s',s}(p_c)$ show interesting scaling behavior whose scaling
exponents could be related to the critical exponents of percolation
\cite{fragperc}, but no phase transition was claimed in the break-up
process. Although
the form of $P_{s',s}(p_c)$ as a function of the daughter mass $s'$
and its dependence on the global size of the fragmenting object $s$
\cite{fragperc} is
similar up to some extent to the behavior of our fragment mass
distribution at the critical point $F(m;\eta_c)$ (see Fig.\
\ref{fig:fig8}), there is no direct analogy between the binary
break-up of percolation clusters and our dynamic fragmentation
process. 

Up to our knowledge no systematic experimental study of the transition
from damage to fragmentation in impact of solids has been performed so
that we cannot 
confront the results of the simulations  with experiments. For a
deeper understanding of the transition predicted by our simulations
further experimental and more analytical theoretical studies are needed.

\section{Acknowledgment}
We thank the referee for drawing our attention to the binary break-up of
percolation clusters.
F.\ Kun is very grateful to  R.\ Englman,  Z.\
J\"ager and K.\ F.\ P\'al for the  valuable 
discussions and for sending him reprints of their works on
fragmentation. F. Kun acknowledges financial support of the Alexander
von Humboldt Foundation (Roman Herzog Fellowship) and that of the projects
SFB381 and OTKA T-023844.

%%%%%%%%%%%%%%%%%%%%%%%%%%%%%%%%%%%%%%%%%%%%%%%%%%%%%%%%%%%%%%%%%%
\begin{table}[h]
\begin{center}
\caption{ The parameter values used in the simulations.}
\begin{tabular}{l c c c c}
$Parameter$ & $Symbol$ & $Unit$ & $Value$  \\
\hline
Failure elongation of a beam & $t_{\epsilon}$ & \% & 3 \\
Failure bending of a beam    & $t_{\Theta}$   & $degree$ & 4 \\
Surface energy & $E_s$ & $erg$ & $1.4 \cdot 10^6$  \\
Damage threshold & $E_d$ & $erg$ & $1.7 \cdot 10^8$ \\
 & $\eta_0$  & 1 & 0.08  \\
Range of velocities & $v_o$ & $\frac{cm}{s}$ & 200 - 3500 \\
\end{tabular}
\label{table_0}
\end{center}
\end{table}

\end{multicols}
\widetext

\newpage
%%%%%%%%%%%%%%%%%%%%%%%%%%%%%%%%%%%%%%%%%%%%%%%%%%%%%%%%%%%%%%%%%%%
\begin{table}[h]
\begin{center}
\caption{Test of the consistency of the critical exponents. 
Starting from the three possible pairs of $\tau, \beta,$ and $\gamma$
the value of the other exponents were derived using the scaling laws.} 
\begin{tabular}{c c c c}
          & $(\tau, \beta) $  & $(\tau, \gamma) $   & $(\beta, \gamma)$  \\
\hline
$\beta$   & $0.11\pm 0.02$ &  $\frac{\gamma \cdot
           (\tau-2)}{3-\tau}=0.09\pm 0.025$ & $ 0.11\pm 0.02$  \\ 
$\gamma$  & $\frac{(3-\tau)\cdot \beta}{\tau-2} = 0.29\pm 0.08$ &
$0.26\pm 0.02$  & $0.26\pm 0.02$ \\ 
$\sigma$  & $\frac{\tau -2}{\beta}=2.45\pm 0.6$  &   
$\frac{3-\tau}{\gamma}=2.8\pm 0.28$ & $\frac{1}{\gamma +
  \beta}=2.7\pm 0.21$  \\
$\tau$    & $2.27\pm 0.05$  &  $2.27\pm 0.05$ & $2 +
\frac{\beta}{\gamma + \beta}=2.3\pm 0.04$ \\
$\alpha$    & $2-\frac{(\tau-1)\cdot\beta}{\tau-2}=1.48\pm 0.12$  &
$2-\frac{(\tau-1)\cdot\gamma}{3-\tau}=1.54\pm 0.06$ &
$2-(2\beta+\gamma)=1.52\pm 0.045$ \\ 
\end{tabular}
\label{table_1}
\end{center}
\end{table}

\begin{multicols}{2} 
\narrowtext
\newpage

%%%%%%%%%%%%%%%%%%%%%%%%%%%%%%%%%%%%%%%%%%%%%%%%%%%%%%%%%%%%%%%%%%

%%%%%%%%%%%%%%%%%%%%%%%%%%%%%%%%%%%%%%%%%%%%%%%%%%%%%%%%%%%%%%%%%%
 
%%%%%%%%%%%%%%%%%%%%%%%%%%%%%%%%%%%%%%%%%%%%%%%%%%%%%%%%%%%%%%%%%%

%%%%%%%%%%%%%%%%%%%%%%%%%%%%%%%%%%%%%%%%%%%%%%%%%%%%%%%%%%%%%%%%%%

%%%%%%%%%%%%%%%%%%%%%%%%%%%%%%%%%%%%%%%%%%%%%%%%%%%%%%%%%%%%%%%%%%

%%%%%%%%%%%%%%%%%%%%%%%%%%%%%%%%%%%%%%%%%%%%%%%%%%%%%%%%%%%%%%%%%%

%%%%%%%%%%%%%%%%%%%%%%%%%%%%%%%%%%%%%%%%%%%%%%%%%%%%%%%%%%%%%%%%%%

%%%%%%%%%%%%%%%%%%%%%%%%%%%%%%%%%%%%%%%%%%%%%%%%%%%%%%%%%%%%%%%%%%

%%%%%%%%%%%%%%%%%%%%%%%%%%%%%%%%%%%%%%%%%%%%%%%%%%%%%%%%%%%%%%%%%%

%%%%%%%%%%%%%%%%%%%%%%%%%%%%%%%%%%%%%%%%%%%%%%%%%%%%%%%%%%%%%%%%%%

%%%%%%%%%%%%%%%%%%%%%%%%%%%%%%%%%%%%%%%%%%%%%%%%%%%%%%%%%%%%%%%%%%

%%%%%%%%%%%%%%%%%%%%%%%%%%%%%%%%%%%%%%%%%%%%%%%%%%%%%%%%%%%%%%%%%%

\end{multicols}
\widetext

\end{document}